\documentclass[]{iopart}
\usepackage{iopams}
\usepackage{setstack}
\usepackage[]{graphicx}
\usepackage{color}
\usepackage[latin1]{inputenc}
\usepackage[english]{babel}

\newcommand{\rP}{{\rm P}}
\newcommand{\If}{{\rm If}}
\newcommand{\rn}{{\rm n}}
\newcommand{\As}{{\rm As}}
\newcommand{\Au}{{\rm Au}}
\newcommand{\La}{{\rm La}}
\newcommand{\rB}{{\rm B}}
\newcommand{\rG}{{\rm G}}

\begin{document}

\title[]{The detection of signals buried in noise}
\author{L Bergamaschi$^1$, G D'Agostino$^1$, L Giordani$^1$, G Mana$^2$, and M Oddone$^3$}
\address{$^1$INRIM -- Istituto Nazionale di Ricerca Metrologica, Unit of Radiochemistry and Spectroscopy,
c/o Department of Chemistry, University of Pavia, via Taramelli 12, 27100 Pavia, Italy}
\address{$^2$INRIM -- Istituto Nazionale di Ricerca Metrologica, str.\ delle Cacce 91, 10135 Torino, Italy}
\address{$^3$Department of Chemistry, University of Pavia, via Taramelli 12, 27100 Pavia, Italy}

\begin{abstract}
This paper examines signal detection in the presence of noise, with a particular emphasis to the nuclear activation analysis. The problem is to decide what between the signal-plus-background and no-signal hypotheses fits better the data and to quantify the relevant signal amplitude or detection limit. Our solution is based on the use of Bayesian inferences to test the different hypotheses.
\end{abstract}

\submitto{Metrologia}

% 06.20.Dk Measurement and error theory
% 02.50.Cw Probability theory
% 07.05.Kf Data analysis: algorithms and implementation; data management
% 02.50.Tt Inference methods

\ead{g.mana@inrim.it}

\section{Introduction}
For signals buried into noise, to decide between the detected and non-detected statements is a long debated problem; in addition, any non-detected decision must include a detection-limit statement. For instance, in analytical chemistry, the detection limit is defined as the lowest quantity of a substance that can be distinguished from no substance at all to within a stated confidence limit \cite{GoldBook}. The orthodox approach to the estimate of detection limits \cite{Currie:1968} is based on the concept of confidence interval and of its interpretation as outlined in seminal papers by Neyman \cite{Neyman:1935,Neyman:1937}. We investigate an alternative approach that uses Bayesian inferences to test the different hypotheses and to quantify the signal amplitude or detection limit.

Using the nuclear activation analysis as an example, that is, the detection of the nuclear activity of a radioisotope in a background photon-count, we illustrate how Bayesian inferences can be used to chose between the signal-plus-background and no-signal hypotheses and to quantify the signal amplitude or detection limit. The contaminant amounts are linked to what it is observed -- the photon numbers in given energy bins -- by calibration factors. The sampling statistics applies to the counts and, therefore, our paper deals only with the observed signal and its associated noise, but the conclusions can be easily extended to the concentrations. As regards as the terminology, the background count is what would be observed in a non-contaminated sample, the gross count is what is actually observed, and the net count is what would be observed in the absence of background. The term measurand will indicate the mean net-count, whereas the terms background- and gross-signal will indicate the expected background- and gross-count.

According the Neyman's view, the detection limits evaluated from the results of a large set of repeated measurement must bound a fixed measurand value with a given frequency. The detection-limit calculation uses hypothesis testing and the distributions of the measurement results given opposite hypotheses. Firstly, the sampling distribution of the background is used to establish a critical limit $L_C$ such that, if the measurand is zero and the count is only noise (null hypothesis), a net count smaller than $L_C$ would be obtained with a high probability, say 95\%. Next, this statement is reversed by choosing the detection limit $L_D$ in such a way that, if the measurand is more than $L_D$ (alternative hypothesis), a net count greater than $L_C$ would be obtained with a high probability, say 95\%.

When the measurand value matters, this frequency-of-occurrence view is not enough. For instance, decisions require probability assignments to propositions that assert the measurand value and, in turns, they require the application of the Bayes theorem \cite{Jaynes,McKay,Gregory,Sivia}. In the Bayesian approach, signal detection and signal estimation are not independent problems and, in a large set of equal measurement results, the detection limit must bound different measurand values with a given frequency. Hypothesis test requires to compare the probability of each hypothesis is true, given the data; hence, the detected or non-detected choice is done according to the maximal probability \cite{Mana:2012}. Only after such a choice has been done representative values -- for example, the mode, mean, or median -- and confidence intervals -- for example, bounding the measurand with a 95\% probability -- can be calculated.

\section{Data model}
Measurements of the impurity concentrations of the $^{28}$Si crystal used for the determination of the Avogadro constant \cite{NA:PRL,NA:Metrologia} are essential to prevent biased results or underestimated uncertainties. The existing literature indicates that Si crystals are extremely pure, but, to obtain a direct evidence of purity, we developed an analytical method based on neutron activation \cite{DAgostino:2012}. Nuclear activation analysis is based on the detection and counting of the $\gamma$ rays emitted by the radioactive isotopes produced by the neutron irradiation. When a neutron is captured by a nucleus, a compound nucleus is formed in an excited state. This step is followed by a prompt de-excitation to a more stable configuration; the new nucleus is usually radioactive and will de-excite by emitting delayed $\gamma$ rays or particles. In the last case the resulting nucleus is often still exited and a further $\gamma$ emission could occur. The energy spectrum of the emitted $\gamma$ rays shows discrete peaks, which identify and quantify the radioactive nuclei and, consequently, the parent contaminants. After calibration against a known amount of contaminant, the number of counts stored in the energy bins relevant to peak gives the impurity contents of the sample.

The gross count $n_\rG$ recorded in a given time interval in any bin of the multichannel analyzer includes a background count $n_\rB$; in addition, owing to the high purity of the Si sample, for almost all the elements, the net count, if any, is deeply buried in the background. To extract all the available information, we assume that the $n_\rG$ and $n_\rB$ data are independent random-numbers. Hence, had the background and gross signals been $\Lambda_\rB$ and $\Lambda_\rG$, their sampling statistics,
\numparts\begin{equation}\label{samplingB}
 \rP_{\rG,\rB}(n_{\rG,\rB}|\Lambda_{\rG,\rB}) = \frac{\Lambda_{\rG,\rB}^{n_{\rG,\rB}} \rme^{-\Lambda_{\rG,\rB}}}{n_{\rG,\rB}!}
\end{equation}
are Poisson distributions having means $\Lambda_\rB$ and $\Lambda_\rG$ and their joint sampling distribution is
\begin{equation}\label{sampling}
 \rP_{\rB\rG}(n_\rB,n_\rG|\Lambda_\rB,\Lambda_\rG) =
 \frac{\Lambda_\rB^{n_\rB} \Lambda_\rG^{n_\rG} \rme^{-(\Lambda_\rB + \Lambda_\rG)} }{n_\rB! n_\rG!} .
\end{equation}\endnumparts
The problem is, firstly, to decide between the detected and non-detected statements and, secondly, to quantify the net signal $\Lambda=\Lambda_\rG-\Lambda_\rB$ or its detection limit.

\begin{figure}
\includegraphics[width=65mm]{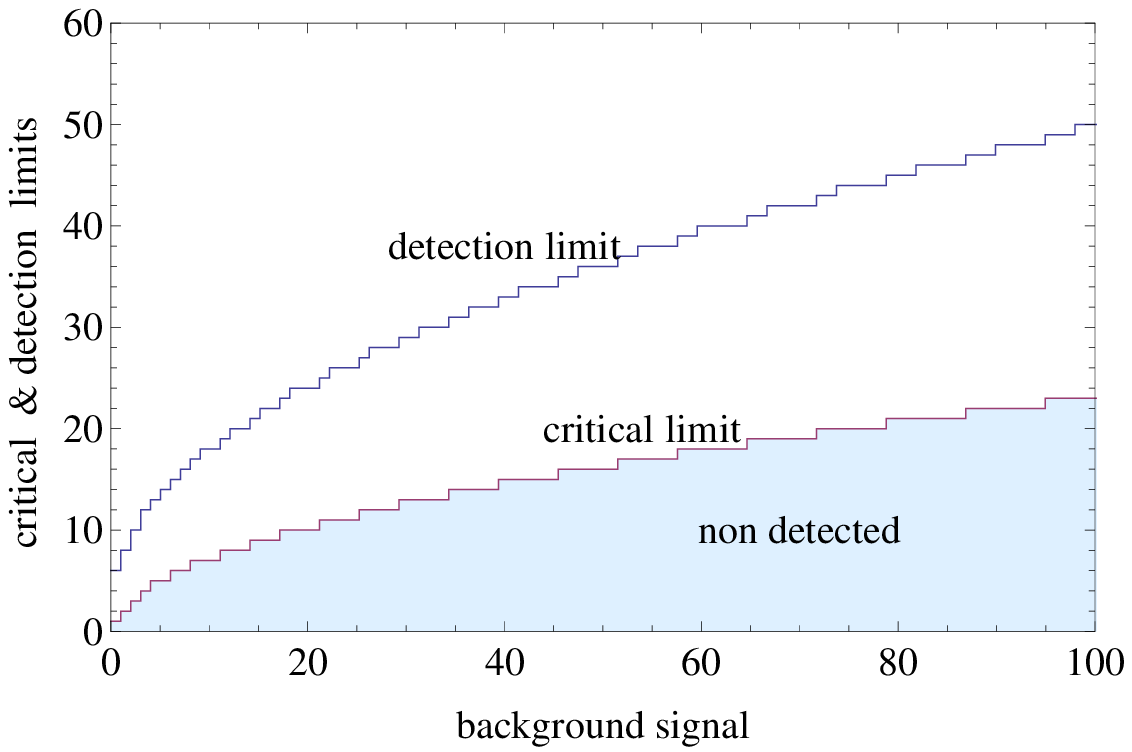}
\includegraphics[width=65mm]{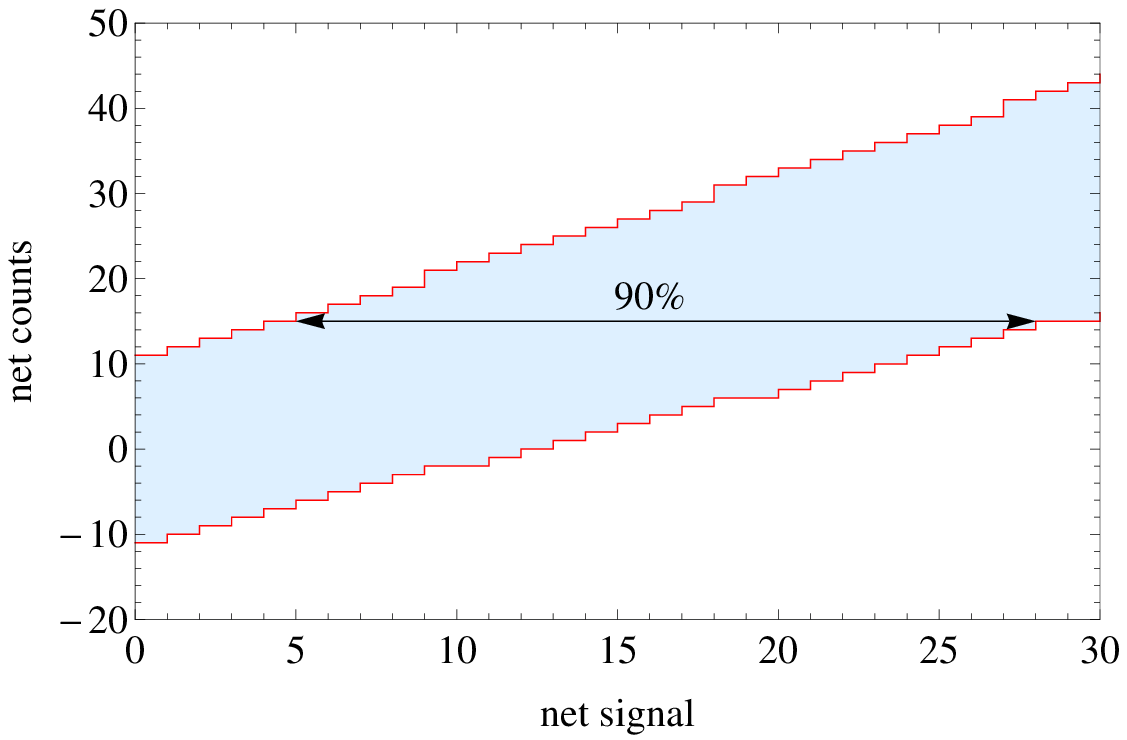}
\caption{Left: 95\% critical limit (lower curve) and 5\% detection limit (upper curve) calculated according to the Currie's construction for a net signal buried in the background noise. The shaded area is the 95\% quantile of the background noise. Right: 5\% (lower line) and 95\% (upper line) quantiles for $n =n_\rG - n_\rB$, when $\Lambda_\rB=20$. The arrow indicates the Neyman's 90\% confidence interval for $\Lambda$ when $n=15$.}\label{interval}
\end{figure}

\section{Classical analysis}
The Currie's construction of the detection limit is as follows \cite{Currie:1968}. The distribution of the minimum-variance unbiased estimate $n = n_\rG - n_\rB$ of $\Lambda$ is the Skellam probability density \cite{Irwin:1937,Skellam:1946}
\begin{equation}\label{Skellam}
 {\rm Pdf}_{\rm Skl}(n|\Lambda_\rG,\Lambda_\rB) = \rme^{-(\Lambda_\rG+\Lambda_\rB)} {\rm I}_{n} \left( 2\sqrt{\Lambda_\rG\Lambda_\rB} \right) \sqrt{\left(\frac{\Lambda_\rG}{\Lambda_\rB}\right)^n} ,
\end{equation}
where ${\rm I}_{n}(x)$  is the modified Bessel function of the first kind and the mean and variance of the net count $n$ are $\langle n \rangle = \Lambda_\rG-\Lambda_\rB=\Lambda$ and $\sigma^2_{n}=\Lambda_\rG+\Lambda_\rB$, respectively. Hence, provided $\Lambda_\rB$ is known -- which is a crucial assumption, the critical limit, $L_C=\lceil x \rceil$, is the smallest integer greater than or equal to the solution of
\begin{equation}\label{CL}
 {\rm Cdf}_{\rm Skl}(x|\Lambda_\rB,\Lambda_\rB) = \alpha ,
\end{equation}
where ${\rm Cdf}_{\rm Skl}$ is the cumulative distribution of ${\rm Pdf}_{\rm Skl}$ and, for instance, $\alpha=0.95$. Therefore, if the net signal is zero and the gross count is only background, a net count greater than $L_C$ would be obtained with a low probability $1-\alpha$. The net signal is assumed detected if $n_\rG > L_C$ and non-detected otherwise. The detection limit, $L_D=x$, is the solution of
\begin{equation}\label{DL}
 {\rm Cdf}_{\rm Skl}(L_C|\Lambda_\rB+x,\Lambda_\rB) = \beta ,
\end{equation}
where, for instance, $\beta=0.05$. It is worth noting that a prior knowledge of the background signal $\Lambda_\rB$ is again assumed. The figure \ref{interval} (left) illustrates the procedure; if the net signal is more than $L_D$, at least the 95\% of the net counts are more than $L_C$.

We can circumvent the need to know $\Lambda_\rB$ in advance by using the Neyman's construction \cite{Neyman:1935,Neyman:1937}; a review can be found in \cite{Feldman:1998}. Actually, this construction produces confidence regions for the $(\Lambda_\rB, \Lambda)$ pair, but, for the sake of simplicity, we fix the value of $\Lambda_\rB$ and calculate a confidence interval for the net signal alone. To this end, following Neyman, we introduce a pair of continuous and monotonic functions of $\Lambda$, $n_1(\Lambda)$ and $n_2(\Lambda)$, so chosen as $[n_1,n_2]$ is an $\alpha$-interval for $n$. That is,
\begin{equation}
 {\rm Cdf}_{\rm Skl}(n_2|\Lambda_\rB+\Lambda,\Lambda_\rB) - {\rm Cdf}_{\rm Skl}(n_1|\Lambda_\rB+\Lambda,\Lambda_\rB) = \alpha .
\end{equation}
Provided the net count is in the domain of the inverse functions $\Lambda_1=n_2^{-1}(n)$ and $\Lambda_2=n_1^{-1}(n)$,
\begin{equation}
 {\rm Prob}\big( \Lambda \in [\Lambda_1,\Lambda_2] | \Lambda \big) = \alpha ,
\end{equation}
by construction and whatever the measurand value may be. Hence, $[\Lambda_1,\Lambda_2]$ is the sough $\alpha$-interval. The figure \ref{interval} (right) illustrates the procedure in the case when $\alpha=0.90$ \cite{Feldman:1998}. According the Neyman's viewpoint, in a long series of repeated measurements with fixed gross $\Lambda_\rG$ and background $\Lambda_\rB$ values, the 90\% of the intervals calculated as indicated by the arrow will contain the measurand $\Lambda=\Lambda_\rG-\Lambda_\rB$.

\subsection{Conceptual limits}
The Currie's constructions of the critical- and detection-limit rely on the prior knowledge of the background signal, which is not available. In practice, the background count $n_\rB$ substitutes for $\Lambda_\rB$, but this does not remove the conceptual difficulty.

An alternative is to use the net count $n=n_\rG-n_\rB$ to determine the Neyman's upper limit of the net signal. However, since the sampling distribution of $n$ depends also on $\Lambda_\rB$, the Neyman's upper limit is still a function of the background signal. Additional troubles arise when $n_\rG<n_\rB$, because the unbiased and minimum-variance estimate of $\Lambda$ is negative and unphysical \cite{Feldman:1998}.

\section{Bayesian analysis}
The problems inherent in the classical analysis can be solved by Bayesian inferences. They are based on the product rule of probabilities
\begin{equation}\label{p-rule}\fl
 \digamma_{\rB\rG}(\Lambda_\rB,\Lambda_\rG|n_\rB,n_\rG) Z_{\rB\rG}(n_\rB,n_\rG) = \rP_{\rB\rG}(n_\rB,n_\rG|\Lambda_\rB,\Lambda_\rG) \pi(\Lambda_\rB,\Lambda_\rG),
\end{equation}
where $\pi(\Lambda_\rB,\Lambda_\rG)$ is the joint probability distribution of the signal values prior the data are available, $\digamma_{\rB\rG}(\Lambda_\rB,\Lambda_\rG|n_\rB,n_\rG)$ is the joint probability distribution of the signal values -- given the signal-plus-background hypothesis and after the data were collected, the likelihood that the signals are $\Lambda_\rB$ and $\Lambda_\rG$ is the sampling distribution $\rP_{\rB\rG}(n_\rB,n_\rG|\Lambda_\rB,\Lambda_\rG)$ evaluated in $n_\rB$ and $n_\rG$, and the evidence of the data model is the probability distribution $Z_{\rB\rG}(n_\rB,n_\rG)$ of the data, no matter what the signals may be.

\subsection{Pre-data distribution}
A key step to calculate $\digamma_{\rB\rG}(\Lambda_\rB,\Lambda_\rG|n_\rB,n_\rG)$ is to assign the density $\pi(\Lambda_\rB,\Lambda_\rG)$ in the $(\Lambda_\rB,\Lambda_\rG)$ points of the signal-value space prior the measurement results are available. In fact, the only way to assign probabilities to the signal values consistent with the measurement results is to update, according the Bayes theorem, the assignments made before the data are at hand. These prior assignments must embed all the in\-for\-mation available, but, to avoid inferences affected by non-available data, no more than this information must be used.

By using the product rule of the probability algebra we can write
\begin{equation}
 \pi(\Lambda_\rB,\Lambda_\rG) = \pi_\rG(\Lambda_\rG|\Lambda_\rB)\pi_\rB(\Lambda_\rB) ,
\end{equation}
where $\pi_\rB(x)$ and $\pi_\rG(x)$ have the same functional form, say, $\pi(x)$, both the signals are strictly positive, and $\Lambda_\rG \ge \Lambda_\rB$. Eventually, $\pi$ must be uninformative. Therefore, we impose scale invariance \cite{Jaynes:1968}. Hence, if $\pi(x)=f(x)$, then $\pi'(kx)=f(kx)$ no matter what the $k$ value may be, where $\pi'(kx)=f(x)/k$ is the probability distribution of $x'=kx$. This ensures that the functional form of $\pi$ is independent of the duration of the counting interval. The reason is that, otherwise, we will embed into $\pi$ -- through a specific $f$-choice -- an information about this duration. Scale invariance limits the $\pi$ choice to the Jeffreys' $\pi(x) \propto 1/x$ distribution.

In the $[0,\infty]$ support, this distribution is not normalizable; therefore, we limit its support to $0 < \Lambda_{\min} < \Lambda_\rB < \Lambda_{\max}$ and $\Lambda_\rB \le \Lambda_\rG < \Lambda_{\max}$ so that
\begin{equation}\fl\label{prior:0}
 \pi(\Lambda_\rB,\Lambda_\rG) = \frac
 {\If(\Lambda_{\min} < \Lambda_\rB < \Lambda_{\max})\If(\Lambda_\rB \le \Lambda_\rG < \Lambda_{\max})}
 {\Lambda_\rB \Lambda_\rG \ln(\Lambda_{\max}/\Lambda_{\min}) \ln(\Lambda_{\max}/\Lambda_\rB) } ,
\end{equation}
where $\If(\Box)$ is one if its argument is true and zero otherwise. Since this distribution does not allow us to calculate analytically the normalization integrals we will found in the following, it will be approximated as
\begin{equation}\fl\label{prior}
 \pi(\Lambda_\rB,\Lambda_\rG) = \frac
 {2\If(\Lambda_{\min} < \Lambda_\rB < \Lambda_{\max})\If(\Lambda_\rB \le \Lambda_\rG < \Lambda_{\max})}
 {\Lambda_\rB \Lambda_\rG \ln^2(\Lambda_{\max}/\Lambda_{\min})} .
\end{equation}
The limits for the support extending from zero to the infinite will be discussed where appropriate. The pre-data distribution of the net signal is the marginal distribution
\begin{eqnarray}\fl \nonumber
 \pi_S(\Lambda) &= &\int_0^\infty \!\!\! \int_{\Lambda_\rB}^\infty \delta(\Lambda_\rG-\Lambda_\rB-\Lambda)
 \pi(\Lambda_\rB,\Lambda_\rG)\, \rmd \Lambda_\rB \rmd \Lambda_\rG \nonumber \\ \fl \label{pre-data}
 &= &\frac
 {2\left[ \ln(\Lambda_{\max}-\Lambda) + \ln(\Lambda_{\min}+\Lambda) - \ln(\Lambda_{\max}\Lambda_{\min}) \right]}
 {\Lambda \ln^2(\Lambda_{\max}/\Lambda_{\min})} ,
\end{eqnarray}
where $0 \le \Lambda < \Lambda_{\max}-\Lambda_{\min}$ and the Dirac delta function $\delta(\Lambda_\rG-\Lambda_\rB-\Lambda)$ is the distribution of $\Lambda$ conditional to the $\Lambda_{\max}$ and $\Lambda_{\min}$ values \cite{Chakraborty:2008}.

\subsection{Post-data distributions}
By combining (\ref{sampling}), (\ref{p-rule}), and (\ref{prior}), the joint probability distribution of the background and gross signals after the data have been collected is
\begin{equation}\fl\label{post}
 \digamma_{\rB\rG}(\Lambda_\rB,\Lambda_\rG|n_\rB,n_\rG) = \frac
 {n_\rB \Lambda_\rB^{n_\rB-1} \Lambda_\rG^{n_\rG-1} \rme^{-(\Lambda_\rB+\Lambda_\rG)}} {(n_\rB+n_\rG-1)! \, _2F_1(n_\rB,n_\rB+n_\rG;n_\rB+1;-1)} ,
\end{equation}
where $_2F_1(a,b;c;z)$ is the hypergeometric function, $Z_{\rB\rG}$ has been obtained by normalization, $[\Lambda_{\min},\Lambda_{\max}]$ has been chosen large enough that the integration limits can be extended from zero to the infinity, $n_\rB > 0$, $n_\rG > 0$, and $\Lambda_\rG \ge \Lambda_\rB > 0$. The post-data distribution of the net signal is the marginal distribution,
\begin{eqnarray}\label{d-marginal}\fl
 \digamma_S(\Lambda|n_\rB,n_\rG) &= &\int_0^\infty \!\!\! \int_{\Lambda_\rB}^\infty \delta(\Lambda_\rG-\Lambda_\rB-\Lambda) \digamma_{\rB\rG}(\Lambda_\rB,\Lambda_\rG|n_\rB,n_\rG)\, \rmd \Lambda_\rB \rmd \Lambda_\rG \nonumber \\
 &= &\int_0^\infty \frac
 {n_\rB \Lambda_\rB^{n_\rB-1} (\Lambda_\rB+\Lambda)^{n_\rG-1} \rme^{-(2\Lambda_\rB+\Lambda)} \, \rmd \Lambda_\rB}
 {(n_\rB+n_\rG-1)! \, _2F_1(n_\rB,n_\rB+n_\rG;n_\rB+1;-1)} \nonumber \\
 &= &\frac{n_\rB \rme^{-\Lambda} \Lambda^{n_\rB+n_\rG-1} {\rm U}(n_\rB, n_\rB+n_\rG+1, 2\Lambda) }{ (n_\rB+n_\rG-1)! \, _2F_1(n_\rB,n_\rB+n_\rG;n_\rB+1;-1)} ,
\end{eqnarray}
where $n_\rB > 0$, $n_\rG > 0$, $\Lambda > 0$, and ${\rm U}(a,b,z)$ is the confluent hypergeometric function. Representative values -- for example, the mode, mean, or median -- and confidence intervals can be calculated from (\ref{d-marginal}), but, contrary to a Neyman's interval, a Bayesian interval is such that, in a long series of repeated measurements of different net signals giving the same net count $n$, a given fraction of the net signals in it.

\subsection{Model selection}
The no-signal hypothesis means that the joint sampling distribution of the data is
\begin{equation}\label{onlyB}
 \rP_{BB}(n_\rG,n_\rB|\Lambda_\rB) =  \frac{ \Lambda_\rB^{n_\rB+n_\rG} \rme^{-2\Lambda_\rB} }{n_\rB! n_\rG!} .
\end{equation}
Consequently, given the no-signal hypothesis, the post-data probability distribution of the background signal is
\begin{equation}\label{post-env}
 \digamma_{BB}(\Lambda_\rB|n_\rG,n_\rB) =
 \frac{\Lambda_\rB^{n_\rB+n_\rG-1} \rme^{-2\Lambda_\rB} }{2^{n_\rB+n_\rG}(n_\rB-1)! (n_\rG-1)! (n_\rB+n_\rG-1)!} ,
\end{equation}
where $Z_{BB}$ has been obtained by normalization, $[\Lambda_{\min},\Lambda_{\max}]$ has been chosen large enough that the integration limits can be extended from zero to the infinity, $\Lambda_\rB > 0$, and $n_\rB+n_\rG>0$.

To chose between the signal-plus-background and no-signal hypotheses, $H_{\rB\rG}$ and $H_{BB}$, that is, between the joint sampling distributions (\ref{sampling}) and (\ref{onlyB}), we need the probability that each hypothesis is true given $n_\rB$ and $n_\rG$ \cite{Mana:2012}. On the assumption that, before the data are available, the probabilities of the two hypotheses are the same, the post-data probabilities of $H_{\rB\rG}$ and $H_{BB}$ are proportional through the same factor to the evidences
\numparts\begin{eqnarray}\fl \nonumber
 Z_{\rB\rG} &= &\int_{\Lambda_{\min}}^{\Lambda_{\max}} \!\!\! \int_{\Lambda_\rB}^{\Lambda_{\max}} \frac
 {2 \Lambda_\rB^{n_\rB-1} \Lambda_\rG^{n_\rG-1} \rme^{-(\Lambda_\rB + \Lambda_\rG)}\, \rmd\Lambda_\rB \rmd\Lambda_\rG}
 {n_\rB! \,n_\rG!\, \ln^2(\Lambda_{\max}/\Lambda_{\min})} \\ \fl \label{HSB}
 &= &\frac{2(n_\rB+n_\rG-1)! \, _2F_1(n_\rB,n_\rB+n_\rG;n_\rB+1;-1)}{n_\rB n_\rB! n_\rG! \ln^2(\Lambda_{\max}/\Lambda_{\min})}
\end{eqnarray}
and
\begin{equation}\fl\label{HB}
 Z_{BB} = \int_{\Lambda_{\min}}^{\Lambda_{\max}} \frac{\Lambda_\rB^{n_\rB+n_\rG-1}\rme^{-2\Lambda_\rB} \, \rmd\Lambda_\rB }
 {n_\rB!\, n_\rG!\, \ln(\Lambda_{\max}/\Lambda_{\min})} =
 \frac{(n_\rB+n_\rG-1)! } {2^{n_\rB+n_\rG} n_\rB!\, n_\rG!\, \ln(\Lambda_{\max}/\Lambda_{\min}) } ,
\end{equation}\endnumparts
where $[\Lambda_{\min},\Lambda_{\max}]$ has been chosen large enough that the integration limits can be extended from zero to the infinity and $n_\rB>0$, $n_\rG >0$. In (\ref{HSB}) and (\ref{HB}), $\ln^2(\Lambda_{\max}/\Lambda_{\min})$ and $\ln(\Lambda_{\max}/\Lambda_{\min})$ are Ockham's penalties for the size of the signal space \cite{Jaynes,McKay,Gregory,Sivia}. Hence,
\numparts\begin{eqnarray}\fl
 {\rm Prob}(H_{\rB\rG}|n_\rB,n_\rG) &= &\frac{Z_{\rB\rG}}{Z_{\rB\rG} + Z_{BB}} \\ \nonumber
 &= &\frac{2^{n_\rB+n_\rG+1}\, _2F_1(n_\rB,n_\rB+n_\rG;n_\rB+1;-1)} {2^{n_\rB+n_\rG+1}\, _2F_1(n_\rB,n_\rB+n_\rG;n_\rB+1;-1) + n_\rB\, \ln(\Lambda_{\max}/\Lambda_{\min})}
\end{eqnarray}
and
\begin{eqnarray}\fl
 {\rm Prob}(H_{BB}|n_\rB,n_\rG) &= &\frac{Z_{BB}}{Z_{\rB\rG} + Z_{BB}} \\ \nonumber
 &= &\frac{n_\rB\, \ln(\Lambda_{\max}/\Lambda_{\min})}
 {2^{n_\rB+n_\rG+1}\, _2F_1(n_\rB,n_\rB+n_\rG;n_\rB+1;-1) + n_\rB\, \ln(\Lambda_{\max}/\Lambda_{\min})} .
\end{eqnarray}\endnumparts

The support of the pre-data distribution must be bounded to a non-null lower limit and a finite upper limit. On the contrary, ${\rm Prob}(H_{BB}|n_\rB,n_\rG)$ tends to one and ${\rm Prob}(H_{\rB\rG}|n_\rB,n_\rG)$ tends to zero. This paradoxical result is caused by the largest parameter space of the $H_{\rB\rG}$ hypothesis and, consequently, its largest Ockham's penalty. This could appear a limitation; however, a $[\Lambda_{\min},\Lambda_{\max}]$ choice can be made on the basis of the background information. In addition, from a numerical viewpoint, the logarithm function maps huge $[\Lambda_{\min},\Lambda_{\max}]$ intervals into negligible Ockham's factors.

\begin{table}
\caption{\label{AsAu} Background and gross counts for the measurements of the amounts of Au, La, and As in a sample of the natural silicon crystal WASO04 by neutron activation analysis. The counting interval was 2 h. The 95\% critical and detection limits have been calculated according to the Currie's constructions with the assumption $\Lambda_\rB=n_\rB$.}
\begin{indented}
\item[]\begin{tabular}{@{}lllrrrrr}
\br
element &reaction &energy &$n_\rB$      &$n_\rG$    &$n$    &$L_C$  &$L_D$ \\
        &         &keV    &counts     &counts   &counts &counts &counts\\
\mr
Au &$^{197}$Au$(\rn,\gamma)$ $^{198}$Au   &411.67   &324 &500 &176   &42  &88 \\
La &$^{139}$La$(\rn,\gamma)$ $^{140}$La   &487.02   &306 &284 &$-22$ &41  &85 \\
As & $^{75}$As$(\rn,\gamma)$ $^{76}$As    &559.10   &296 &311 &15    &40  &84 \\
\br
\end{tabular}
\end{indented}
\end{table}

\section{Application example}
\begin{figure}
\includegraphics[width=65mm]{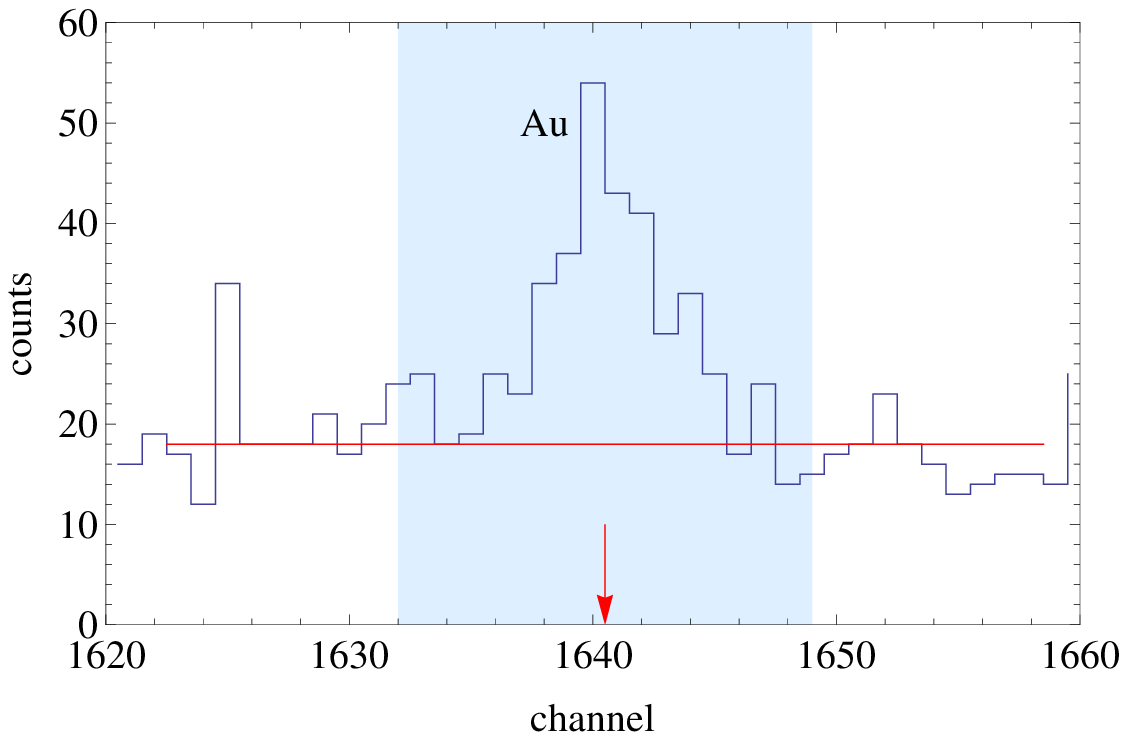}
\includegraphics[width=65mm]{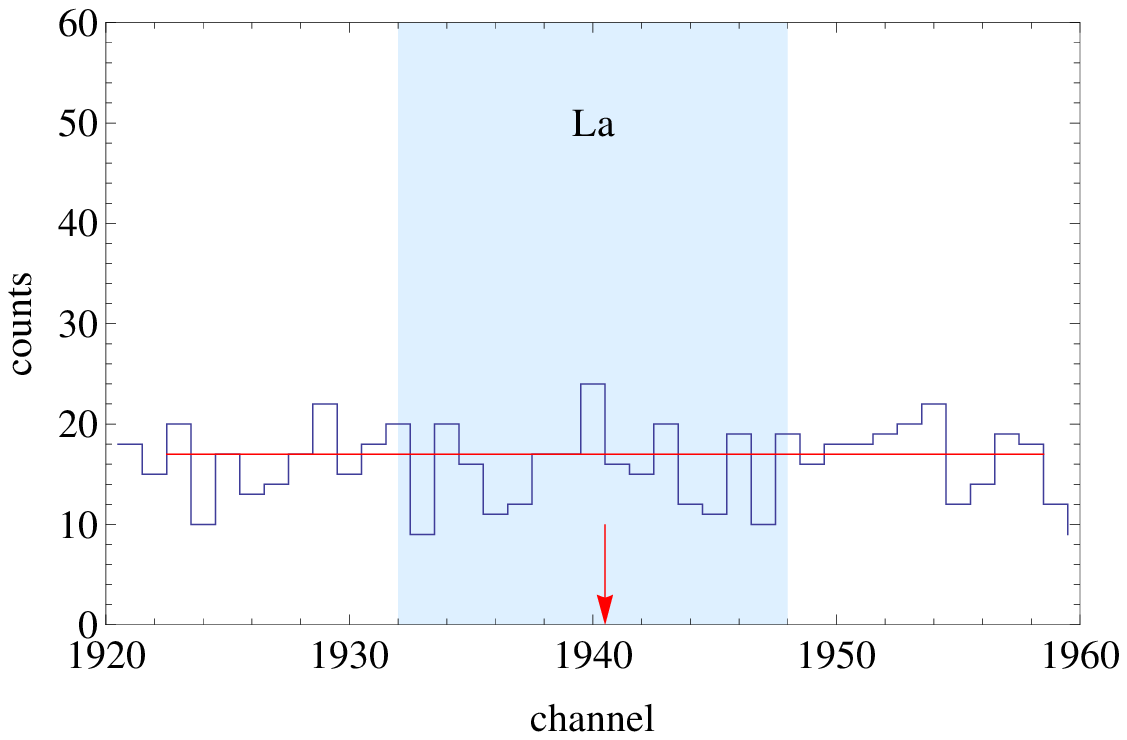}
\includegraphics[width=65mm]{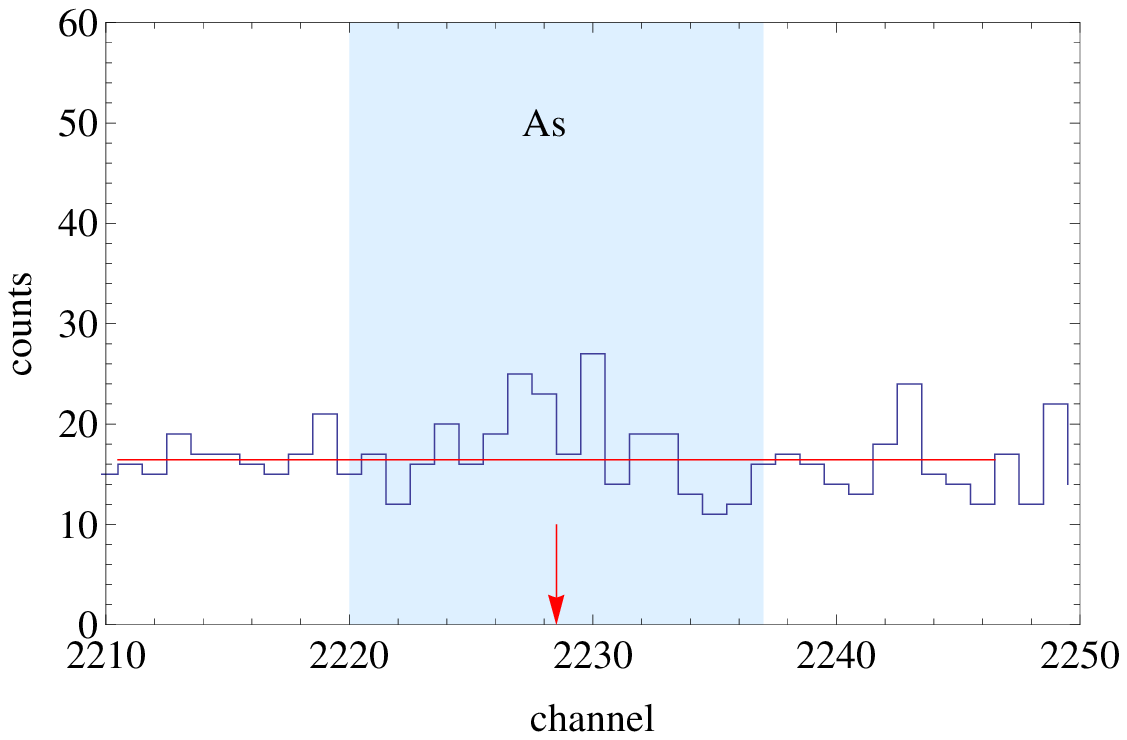}
\caption{Zooms of the emission spectra in the neighbours of the channels (indicated by the arrows) corresponding to the energies of the $\gamma$ rays emitted in the de-excitation of the activated Au, La, and As nuclei. The shaded areas indicate the peak widths. The horizontal lines indicate the background counts, as estimated from the peak tails; the line lengths indicate the tail-channels considered.}\label{spectra}
\end{figure}

As an application example, we consider the measurements of the amounts of Au, La, and As in a sample of the natural silicon crystal WASO04 by neutron activation analysis \cite{DAgostino:2012}. Zooms of the emission spectra in the neighbours of the channels corresponding to the energies of the $\gamma$ rays emitted in the de-excitation of the activated nuclei are shown in Fig.\ \ref{spectra}. All the photons collected in the bins included in each peak (chosen as five times the calibrated full peak half width) were added to obtain the gross counts. The background counts were estimated by adding all the photon collected in an equal number of tail channels fairly subdivided between in the left and right tails. The relevant reactions, peak energies, and background, gross, and net counts are given in table \ref{AsAu}. The 95\% critical and detection limits have been calculated according to the Currie's constructions (\ref{CL}) and (\ref{DL}); they are shown in table \ref{AsAu}. Their meanings are as follow: if the net signal is zero, the probability that the net count is less than $L_C$ is 0.95; if the net signal is more than $L_D$, the probability that the net count is more than $L_C$ is 0.95. Accordingly, only a gold contamination has been detected.

\begin{figure}[b]
\includegraphics[width=67mm]{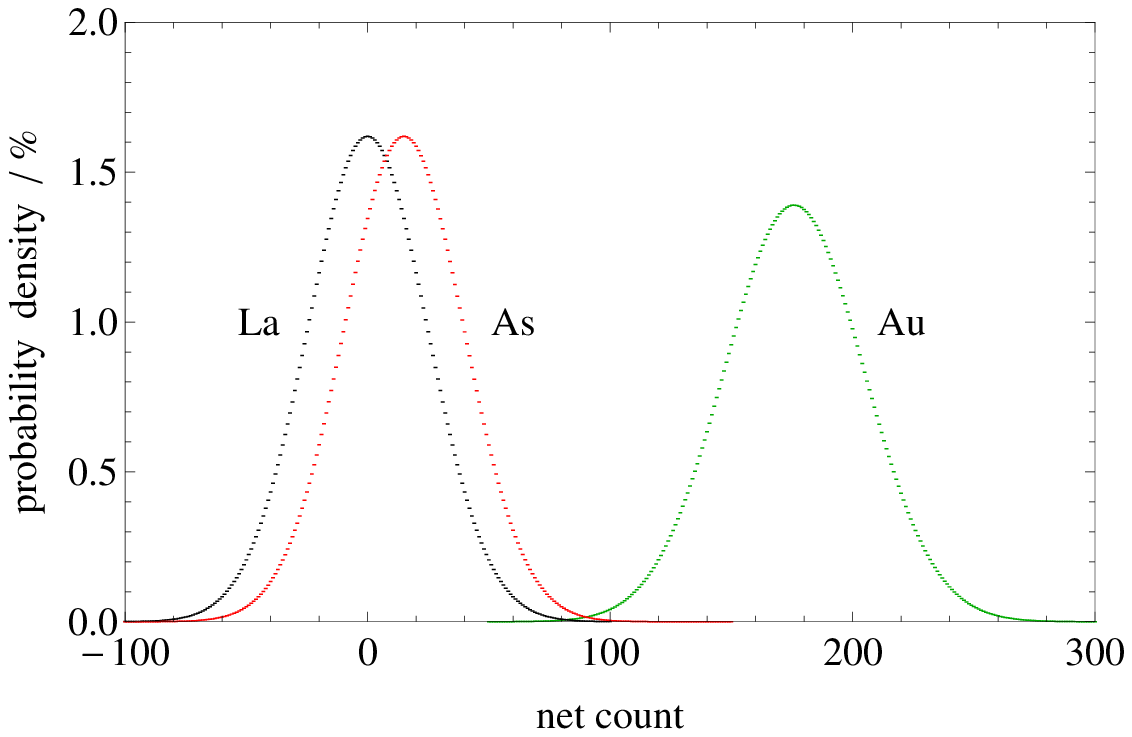}
\includegraphics[width=65mm]{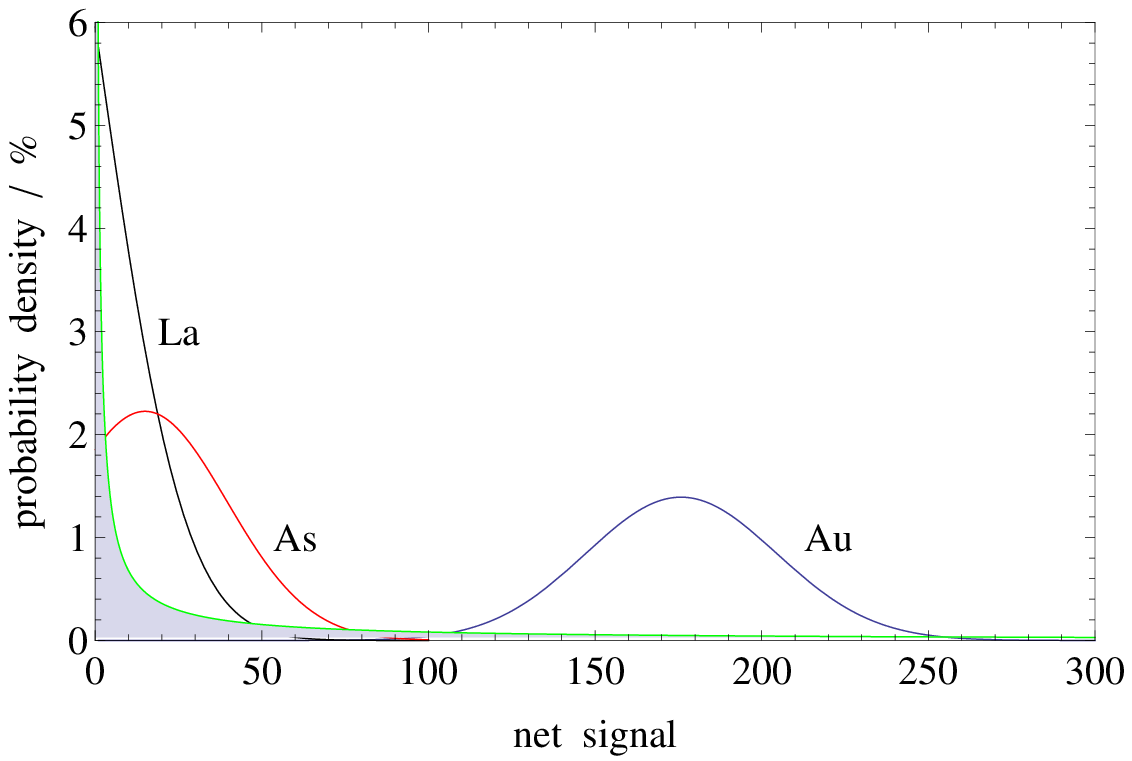}
\caption{Left: sampling-distributions of the unbiased minimum-variance estimates of the net signal for Au, La, and As. Right: post-data distributions of the net signals. The filled curve is the pre-data distribution (\ref{pre-data}).}\label{marginal}
\end{figure}

The unbiased  minimum-variance estimates of the gold, lanthanum, and arsenic net-signals are $n(\Au)=n_\rG(\Au)-n_\rB(\Au)$, $n(\La)=n_\rG(\La)-n_\rB(\La)$, and $n(\As)=n_\rG(\As)-n_\rB(\As)$; they are given in table \ref{results} together with the relevant standard deviations. The standard deviations have been calculated by using the Skellam distribution (\ref{Skellam}), where the  estimates $n_\rB$ and $n_\rG$ of the background and gross signals have been used. The hypothetical Skellam sampling-distributions of the net counts are shown in Fig.\ \ref{marginal} (left). To calculate the actual sampling distributions would require knowing the the background- and gross-signal values in advance; in Fig.\ \ref{marginal}, they were set equal to the background- and gross-signal counts with the exception of the $n(\La)$ distribution, where both were set equal to the $[n_\rG(\La)+n_\rB(\La)]/2$ mean. It is worth noting that $n(\La)$, though a perfectly legitime unbiased estimate of $\Lambda(\La)$, is negative and non-physical. Table \ref{results} gives also the 95\% Neyman upper-limits of the net signals, which have been calculated for $\Lambda_\rB=n_\rB$. Their meaning is as follow: in a large set of measurement repetitions, 95\% of upper limits so calculated are more than the net signal. In this frequency-of-occurrence sense, the probability that the net signal is less than the Neyman upper-limits is 0.95.

\begin{figure}
\includegraphics[width=65mm]{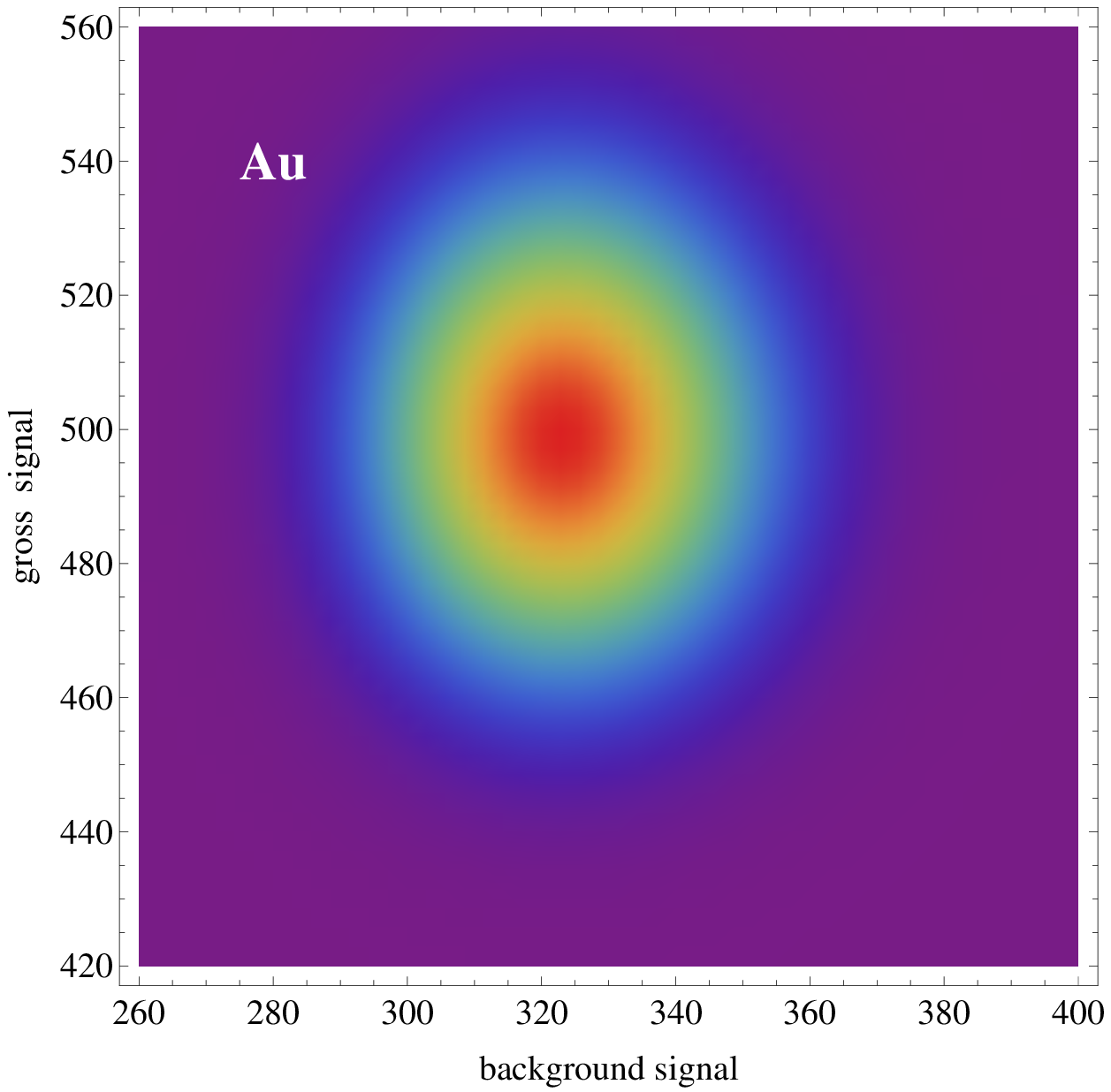}
\includegraphics[width=65mm]{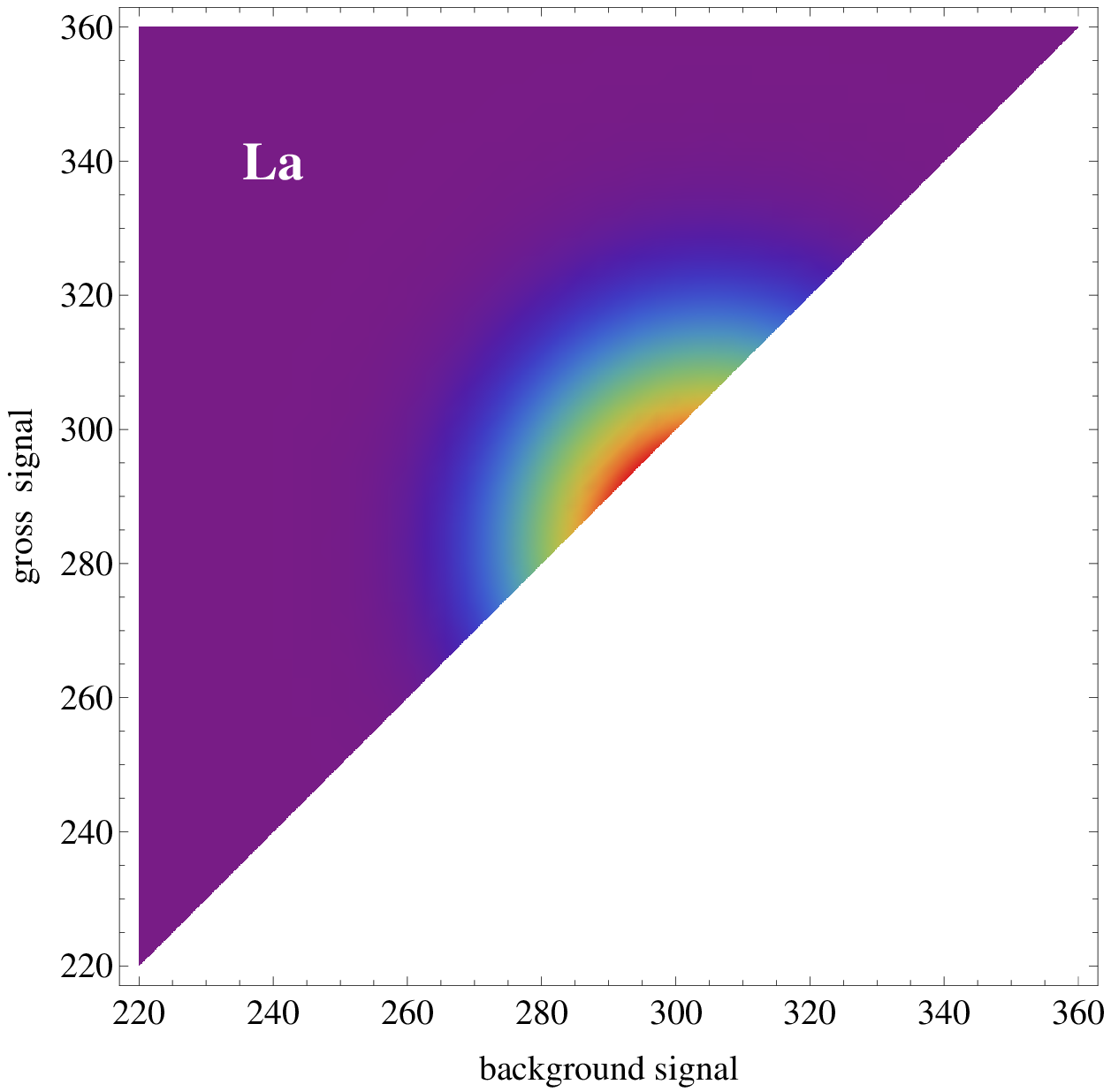}
\includegraphics[width=65mm]{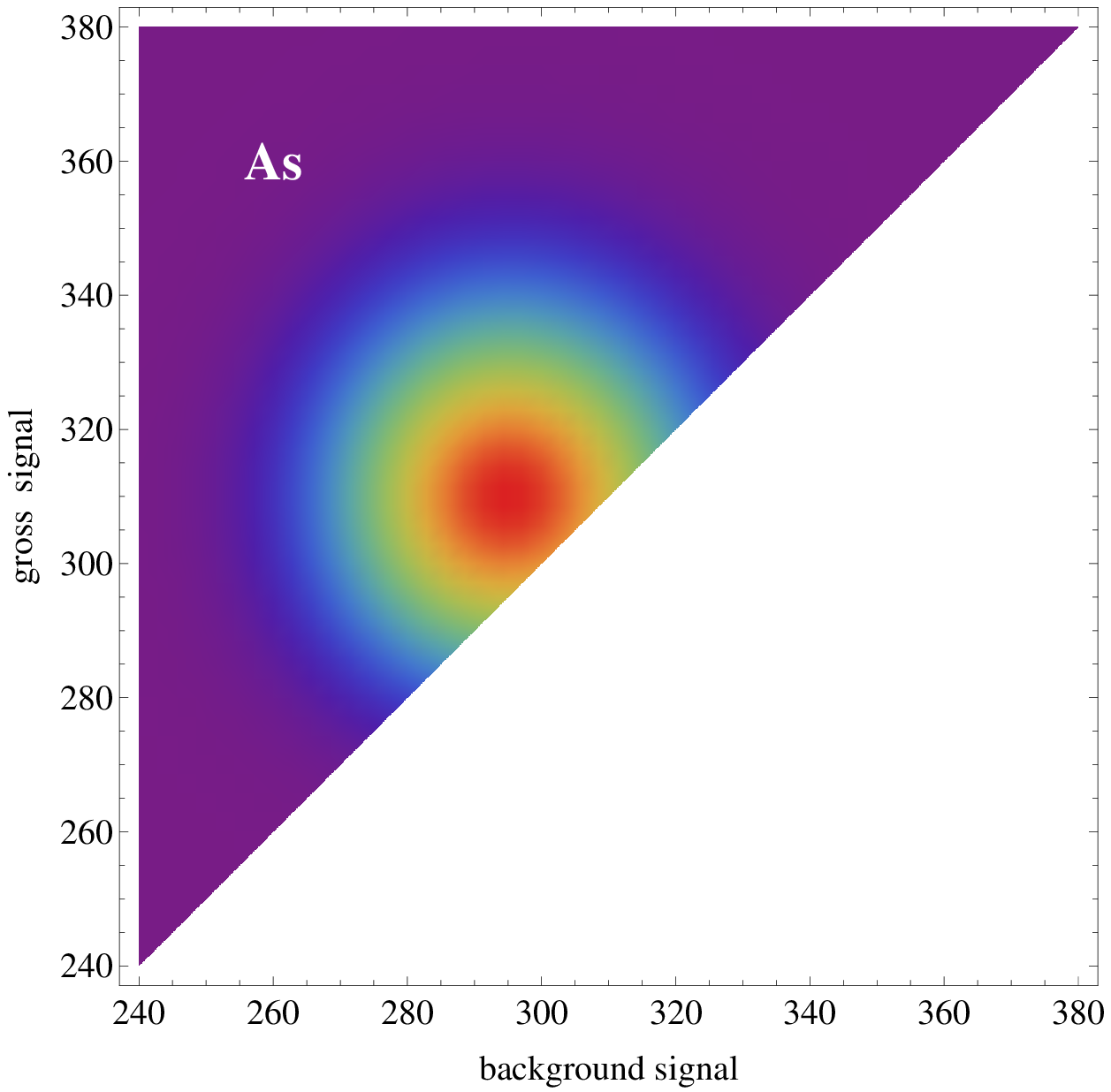}
\caption{Contour plots of the joint post-data distributions of the background and gross signals for Au, La, and As. The white areas are excluded by the prior information $\Lambda_\rG \ge \Lambda_\rB$.}\label{contour}
\end{figure}

The Bayesian joint post-data distributions of the background and gross signals are shown in Fig.\ \ref{contour}, where the support of the pre-data distribution is from $\Lambda_{\min}=10^{-4}$ to $\Lambda_{\max}=10^4$. The relevant marginal distributions of the net signals are given in Fig.\ \ref{marginal} (right). For a comparison, Fig.\ \ref{marginal} shows also the pre-data distribution (\ref{pre-data}).

The probabilities of both the detected and non-detected statements have been calculated according to the evidence of the relevant data models; the results are given in table \ref{evidences}. The gold contamination is evident; the lanthanum and arsenic contamination are very uncertain. Table \ref{results} gives the median of the possible net-signals values together with the 25\% and 75\% quantiles. This table gives also the 95\% Bayesian upper-limits of the net signals, whose meaning is as follow: in a large set of measurement repetitions giving the same background and gross counts, 95\% of the net signals (in principle, different) are less than the limit so calculated. It must be noted that the Bayesian median of the possible $\Lambda(\La)$ values is positive; further discussions of the Bayesian inference of a positive quantity from a negative measurement result can be found in \cite{Calonico:2009a,Calonico:2009b}.

\begin{table}
\caption{\label{results} Bayesian inferences: median net-signals and 95\% upper limits for the Au, La, and As fractions in the WASO04 sample; the sub and super scripts indicate the first and third quartiles. Frequency-of-occurrence viewpoint: unbiased estimates (in parentheses are the standard-deviations) and 95\% Neyman upper limits of the net signals. The Neyman upper limits have been calculated by assuming that $n_\rB=\Lambda_\rB$.}
\begin{indented}
\item[]\begin{tabular}{@{}lrrrr}
\br
 &\multicolumn{2}{c}{Bayesian inferences} &\multicolumn{2}{c}{frequency-of-occurrence} \\
element &median   &95\% interval &$n$               &95\% interval \\
        &counts   &counts        &counts            &counts        \\
\mr
Au &$176_{-19}^{+19}$  &$<223$       &176(29)    &$<225$ \\
La &$10_{-6}^{+9}\,\,$ &$< 35$       &$-22(24)$  &$<20$ \\
As &$24_{-11}^{+14}$   &$< 59$       &15(25)     &$<57$  \\
\br
\end{tabular}
\end{indented}
\end{table}

\begin{table}[b]
\caption{\label{evidences} Evidences and probabilities of the detected and non-detected hypothesis. The support of the prior distribution of the background and gross signals is from $\Lambda_{\min}=10^{-4}$ to $\Lambda_{\max}=10^4$.}
\begin{indented}
\item[]\begin{tabular}{@{}llrlrlr}
\br
hypothesis &\multicolumn{2}{c}{evidence \hspace{1mm} probability} &\multicolumn{2}{c}{evidence \hspace{1mm} probability}
&\multicolumn{2}{c}{evidence \hspace{1mm} probability} \\
\mr
&\multicolumn{2}{c}{Au} &\multicolumn{2}{c}{La} &\multicolumn{2}{c}{As} \\
\mr
detected      &$3.6\times 10^{-8}$  &100\%  &$1.2\times 10^{-8}$ & 1\%  &$4.7\times 10^{-8}$ & 2\% \\
non-detected  &$1.1\times 10^{-14}$ &0\%    &$2.0\times 10^{-6}$ &99\%  &$2.4\times 10^{-6}$ &98\% \\
\br
\end{tabular}
\end{indented}
\end{table}

Nevertheless their different conceptual meanings -- the median of the net-signal value-space and a net-signal measure drawn from an unbiased minimum-variance population of net-signal estimates -- Bayesian estimate and frequency-of-occurrence measure of $\Lambda(\Au)$ are numerically the same. The same is true for the relevant Bayesian and Neyman' confidence intervals, though the first refers to an ensemble of different net-signal values but the same background and gross counts and the second refers to an ensemble of different intervals calculated from different background and gross counts but the same net-signal value. The reason is that both the approaches rely on similar, quasi-Gaussian, probability distributions and that the prior information was irrelevant. Contrary, significant differences are evident when the net count approaches zero or it is negative.

\section{Conclusions}
We showed that probability calculus and Bayesian inferences offers a solution to the problem of deciding between the signal-plus-background and no-signal hypotheses, when looking for quantities whose magnitude is comparable with the background noise of the measurement procedure. Given the measurement results, having been calculated the probabilities of the detected and non-detected hypotheses, optimal decisions follow. For instance, having the signal-plus-background model been selected, a measurand value can be optimally chosen according to the post-data probabilities of its possible values. As regards the detection-limit estimate, the Neyman approach focuses attention on the data processing and it is concerned in finding a statistics capable of a pre-determined performances in the set of the results of repeated measurements of the same measurand. The Bayesian approach -- which focuses attention on the measurand-value probabilities -- is concerned in the set of different measurand values consistent with repeated measurements giving the same result.

\ack
This work was jointly funded by the European Metrology Research Programme (EMRP) participating countries within the European Association of National Metrology Institutes (EURAMET) and the European Union.


\section*{References}
\begin{thebibliography}{99}
\bibitem{GoldBook}
 McNaught A D and Wilkinson A 1997 IUPAC Compendium of Chemical Terminology, 2nd ed.\ (Blackwell Scientific Publications, Oxford)
\bibitem{Currie:1968}
 Currie L A 1968 Limits for Qualitative Detection and Quantitative Determination -- Application to Radiochemistry {\it Analytical Chemistry} {\bf 40} 586-93
\bibitem{Neyman:1935}
 Neyman J 1935 On the problem of confidence intervals {\it Ann.\ Math.\ Stat.} {\bf 6} 111-6
\bibitem{Neyman:1937}
 Neyman J 1937 Outline of a theory of statistical estimation based on the classical theory of probability {\it Philos.\ Trans.\ Roy.\ Soc.\ Ser.\ A} {\bf 236} 333-80
\bibitem{Jaynes}
 Jaynes E T 2003 Probability theory: The logic of science (Cambridge: Cambridge University Press)
\bibitem{McKay}
 Mc Kay D JC 2003 Information Theory, Inference, and Learning Algorithms (Cambridge: Cambridge University Press)
\bibitem{Gregory}
 Gregory P C 2005 Bayesian Logical Data Analysis for the Physical Sciences (Cambridge: Cambridge University Press)
\bibitem{Sivia}
 Sivia D and Skilling J 2006 Data Analysis: A Bayesian Tutorial (Oxford: Oxford University Press)
\bibitem{Mana:2012}
 Mana G, Massa E, and Predescu M 2012 Model selection in the average of inconsistent data: an analysis of the measured Planck-constant values {\it Metrologia} {\bf 49} 492-500
\bibitem{NA:PRL}
 Andreas B {\it et al.} 2011 Determination of the Avogadro constant by counting the atoms in a $^{28}$Si crystal {\it Phys.\ Rev.\ Lett.} {\bf 106} 030801
\bibitem{NA:Metrologia}
 Andreas B {\it et al.} 2011 Counting the atoms in a $^{28}$Si crystal for a new kilogram definition {\it Metrologia} {\bf 48} S1-13
\bibitem{DAgostino:2012}
 D'Agostino G, Bergamaschi L, Giordani L, Mana G, Massa E, and Oddone M 2012 Elemental characterization of the Avogadro silicon crystal WASO 04 by neutron activation analysis {\it Metrologia} {\bf 49} 696-701
\bibitem{Feldman:1998}
 Feldman G J and Cousins R D 1998 Unified approach to the classical statistical analysis of small signals {\it Phys.\ Rev.\ D} {\bf 57} 3873-89
\bibitem{Irwin:1937}
 Irwin J O 1937 The frequency distribution of the difference between two independent variates following the same Poisson distribution {\it J.\ R.\ Stat.\ Soc.\ A} {\bf 100} 415-16
\bibitem{Skellam:1946}
 Skellam  J G 1946 The frequency distribution of the difference between two Poisson variates belonging to different populations {\it J.\ R.\ Stat.\ Soc.\ A} {\bf 109} 296-6
\bibitem{Jaynes:1968}
 Jaynes E T 1968 Prior Probabilities {\it IEEE Trans.\ Sys.\ Sci.\ Cybernetics} {\bf 4} 227-41
\bibitem{Chakraborty:2008}
 Chakraborty S 2008 Some Applications of Dirac's Delta Function in Statistics for More Than One Random Variable {\it Appl.\ Math.} {\bf 3} 42-54
\bibitem{Calonico:2009a}
 Calonico D, Levi F, Lorini L and G Mana 2009 Bayesian inference of a negative quantity from positive measurement results {\it Metrologia} {\bf 46} 267-71
\bibitem{Calonico:2009b}
 Calonico D, Levi F, Lorini L and G Mana 2009 Bayesian estimate of the zero-density frequency of a Cs fountain {\it Metrologia} {\bf 46} 629-36
\end{thebibliography}
\end{document}